\documentclass[prd,aps,preprintnumbers,showpacs,10pt]{revtex4}
\usepackage{graphicx}
\usepackage{epstopdf}
\usepackage{color}
\usepackage{epsfig}
\usepackage{subfigure}

\def\slash#1{\setbox0=\hbox{$#1$}#1\hskip-\wd0\dimen0=5pt\advance
\dimen0 by-\ht0\advance\dimen0 by\dp0\lower0.5\dimen0\hbox
to\wd0{\hss\sl/\/\hss}} 
\begin{document}
\preprint{BARI-TH/597-08}
\title{$\eta_b$ and $\eta_c$ radiative decays in the Salpeter model with the AdS/QCD inspired potential}
\author{\textbf{Floriana Giannuzzi}}

\affiliation{Universit\`a degli Studi di Bari, I-70126 Bari, Italy\\ INFN, Sezione di Bari, I-70126 Bari, Italy}
\begin{abstract}
The decay constants and the radiative decay widths of $\eta_b(nS)$ and $\eta_c(nS)$ are computed within a semirelativistic quark model, using a potential found through the AdS/QCD correspondence. For $\eta_c$, the results are in agreement with experimental data, while in the case of $\eta'_c$ 
 a discrepancy is found and the possible reasons are discussed. 

\end{abstract}

 \maketitle


The recent observation of the bottomonium ground state $\eta_b(1S)$ by the BaBar Collaboration, in the $\Upsilon$(3S) radiative decay mode: $\Upsilon(3S)\rightarrow \gamma \eta_b$ \cite{:2008vj}, has led to a new interest in this particle and its decays. The measured mass is
\begin{equation}
M_{\eta_b}=9388.9^{+3.1}_{-2.3} \mbox{(stat)}\pm2.7\mbox{(syst)  MeV} 
\end{equation} 
corresponding to a hyperfine splitting $M_{\Upsilon(1S)}-M_{\eta_b(1S)}=71.4^{+2.3}_{-3.1} \mbox{(stat)}\pm2.7 \mbox{(syst)}$ MeV \cite{:2008vj}.

On the other hand, the subsequent decay of $\eta_b$ has not been observed. The meson $\eta_b$ is expected to decay mainly to hadrons; other possible modes are the radiative transitions. 

Motivated by this new experimental observation, in this paper a calculation  of the $\eta_b$ decay constant and of the width relative to the decay $\eta_b\rightarrow \gamma\gamma$  is presented, in the framework of the model proposed in Ref. \cite{Carlucci:2007um}.
 Moreover, the calculation has been extended to the radial excitations and to the charmonium corresponding states, since the model can be properly employed for   heavy quarkonium states. 
 This can be useful, since in many  cases the experimental values are not known and since there are some discrepancies among the predictions of different theoretical models and the experimental values. For example, there is only one measurement, by the CLEO   Collaboration, of the  $\eta'_c\rightarrow \gamma\gamma$ radiative decay width:  $\Gamma_{\gamma\gamma}(\eta'_c)=1.3\pm0.6$ KeV  \cite{Asner:2003wv}, a result not reproduced by most of theoretical predictions which suggest larger values.
 Within this model,  the decay constants and the leptonic decay widths of vector $b\bar{b}$ and $c\bar{c}$ mesons are also evaluated and a  comparison  with the experimental results is carried out. 
 
In  the model introduced in ref. \cite{Carlucci:2007um} the meson spectrum is computed solving a semirelativistic wave equation, the Salpeter equation:

\begin{equation}\label{salpeter}
\left(\sqrt{m_1^2-\nabla^2}+\sqrt{m_2^2-\nabla^2}+V(r)\right)
\psi({\bf r})\,=\,M\, \psi({\bf r})\, ,
\end{equation}
where $m_1$ and $m_2$ are the masses of the constituent quarks and $M$ and $\psi({\bf r})$ are the mass and the wave function of the meson, respectively. The $\ell$=0 case is considered. The potential $V(r)$ comprises three terms:
\begin{equation}\label{potenziale}V(r)=V_{AdS/QCD}(r)+V_{spin}(r)+V_0 \,.\end{equation}

\begin{figure}[ht]
\begin{center}
\includegraphics[scale=0.45]{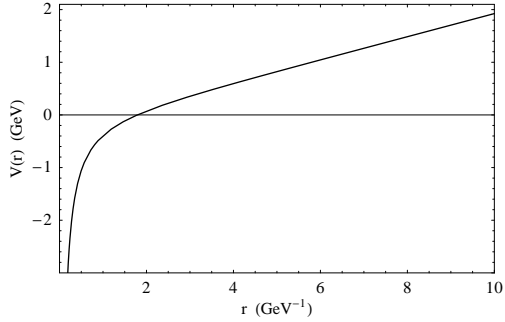}~~~~~~~~~~~~~~~~~~~~~~~~~~~
\end{center}
\caption{\label{pot} The $q\bar{q}$ potential $V_{AdS/QCD}(r)$ obtained from the AdS/QCD correspondence, with $c=0.4$ GeV$^2$ and $g=2.50$. }
\end{figure}

The main feature of the model is that the static potential  $V_{AdS/QCD}(r)$  is obtained evaluating the expectation value of the Wilson loop in the  AdS/QCD framework \cite{Andreev:2006ct}, which provides a holographic model able to describe some aspects of QCD, namely linear confinement,  Regge trajectories, glueball spectrum, the light meson spectrum and decay constants \cite{Karch:2006pv}. The static  $q\bar{q}$ potential is obtained, in this framework,  as a parametric function \cite{Andreev:2006ct}:
\begin{equation}\left\{
\begin{array}{cc}\label{zakpot} 
V_{AdS/QCD}(\lambda)\,=\,\frac{g}{\pi}
\sqrt{\frac{c}{\lambda}} \left( -1+\int_0^1 dv \, v^{-2} \left[
\mbox{e}^{\lambda v^2/2} \left(1-v^4
\mbox{e}^{\lambda(1-v^2)}\right)^{-1/2}-1\right]\right) & \\
r(\lambda)\,=\,2\, \sqrt{\frac{\lambda}{c}} \int_0^1 dv\, v^{2}
\mbox{e}^{\lambda (1-v^2)/2} \left(1-v^4
\mbox{e}^{\lambda(1-v^2)}\right)^{-1/2}  \hspace{3.3cm}&  \,,
\end{array}
\right.\end{equation}
where $r$ is the interquark distance and $\lambda$ varies in the range: $0\leq\lambda<2$. The potential $V_{AdS/QCD}(r)$, therefore, depends on
two parameters, $g$ and $c$; it is depicted in Fig. \ref{pot} for the two values of $c$ and $g$ employed in the present analysis.

The term of the potential $V(r)$ in Eq. (\ref{potenziale}) accounting for the spin-interaction is given by
\begin{equation}V_{spin}(r)\,=\,A \frac{\tilde\delta(r)}{m_1 m_2}{\bf S_1}\cdot{\bf
S_2} \qquad\quad\mbox{with }\qquad \tilde\delta(r)=\left(\frac{\sigma}{\sqrt{\pi}}\right)^3
e^{-\sigma^2 r^2}\,, \end{equation}
and involves the parameters $\sigma$, together with $A_b$ and $A_c$ for the cases of beauty and charm, respectively. The constant term $V_0$ is the last parameter fixing the potential in the Salpeter equation.

The singularity of the wave function at $r=0$ is regulated introducing a value $r_{min}$, so that at smaller distances the potential is constant and equal to $V(r_{min})$. 
At odds with the analysis in \cite{Carlucci:2007um}, where $r_{min}$ is an additional input parameter in the fit of the meson spectrum, here we fix $r_{min}$ according to a QCD duality argument \cite{Cea:1986bj}: $r_{min}=\frac{4\pi}{3 M}$, where $M$ is the mass of the meson, so that the input set of parameters includes $c=$0.4 GeV$^2$, $g$=2.50 and $V_0$=-0.47 GeV for the AdS/QCD potential, $A_c$=14.56, $A_b$=6.49, and  $\sigma$=0.47 GeV for the spin term, and the constituent quark masses  $m_c$=1.59 GeV and $m_b$=5.02 GeV; the values of all the parameters are fixed by a best fit of the meson masses computed in the model to the observed heavy meson masses \cite{pdg}.

Within the Salpeter model, the decay constants $f_P$ and $f_V$ of a pseudoscalar and a vector meson, defined  by 
\begin{eqnarray}\label{defdec}
\langle0|A^\mu_{ij} |P(k)\rangle&=&i k^\mu Q_{ij} f_P  \nonumber \\
\langle0|V^\mu_{ij} |V(k,\lambda)\rangle&=& \epsilon(\lambda)^\mu Q_{ij} m_V f_V 
\end{eqnarray}
respectively, where $k$ is the momentum, $\lambda$ the helicity and $\epsilon$ the polarization vector of the meson, are given by \cite{Colangelo:1990rv}
\begin{eqnarray}\label{decayconst}
f_P&=&\sqrt{3} \frac{1}{2\pi M} \int_0^{+\infty} dk \; k \; \tilde{u}_0(k) N^{\frac{1}{2}} \left[ 1-\frac{k^2}{(E_i+m_i)(E_j+m_j)} \right] \nonumber\\
f_V&=&   \sqrt{3} \frac{1}{2\pi M} \int_0^{+\infty} dk \; k \; \tilde{u}_0(k) N^{\frac{1}{2}} \left[ 1+\frac{k^2}{3(E_i+m_i)(E_j+m_j)} \right]
\end{eqnarray}
with
$$N=\frac{(E_i+m_i)(E_j+m_j)}{E_j E_i} \, .$$
In (\ref{defdec}), $A^\mu_{ij}$ is the axial current $\bar{q}_i \gamma_5\gamma^\mu q_j$,   $V^\mu_{ij}$ is the vector current $\bar{q}_i \gamma^\mu q_j$ and $Q_{ij}$ is the meson flavor matrix. In (\ref{decayconst}), $M$ is the mass of the meson, $m_i$ is the mass of the constituent quark $i$ and $E_i$ its energy, $\tilde{u}(k)$ is the meson reduced wave function  in momentum space, obtained by Fourier transforming  the reduced radial wave function $u(r)=r \, \psi(r)$; and $k$ is the momentum of the constituent quark in the rest frame of the meson.  

\begin{figure}[h]
\subfigure{
\includegraphics[scale=0.48]{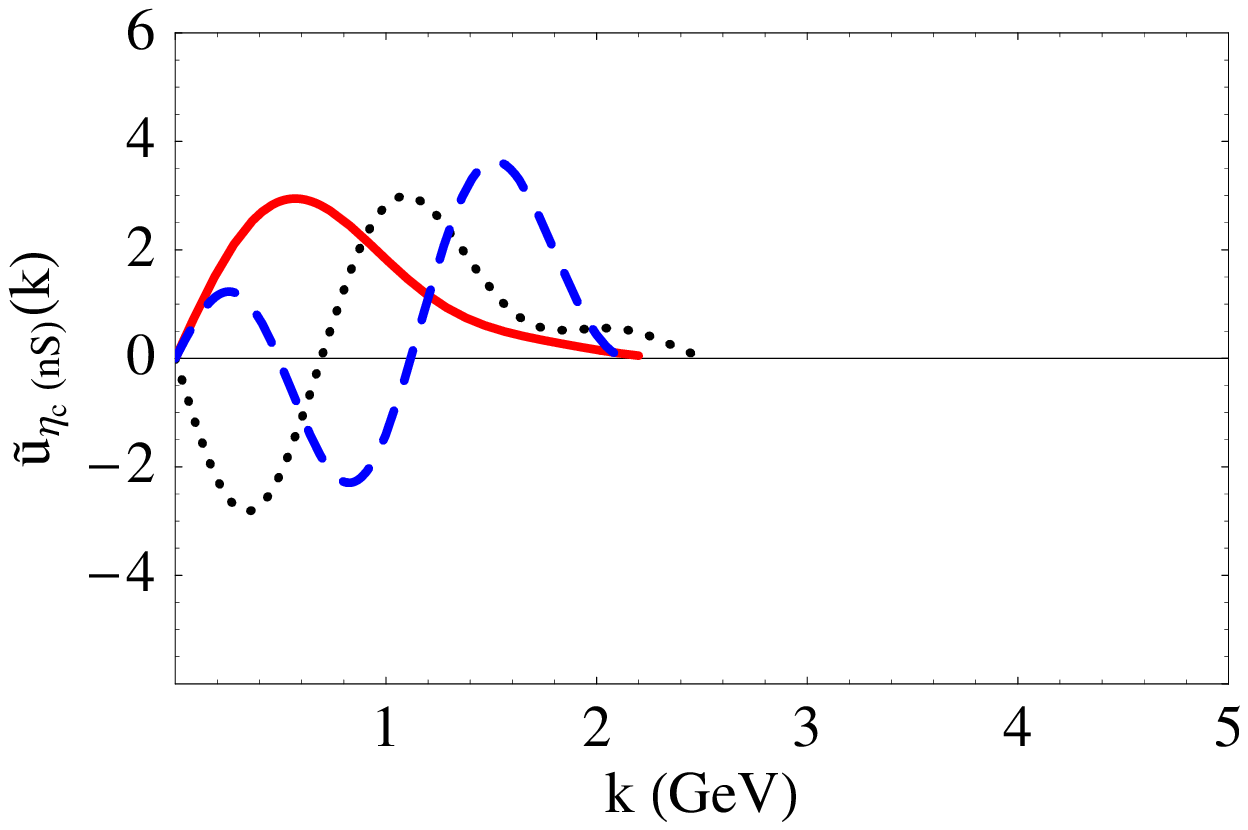}
}
\hspace*{1cm}
\subfigure{
\includegraphics[scale=0.48]{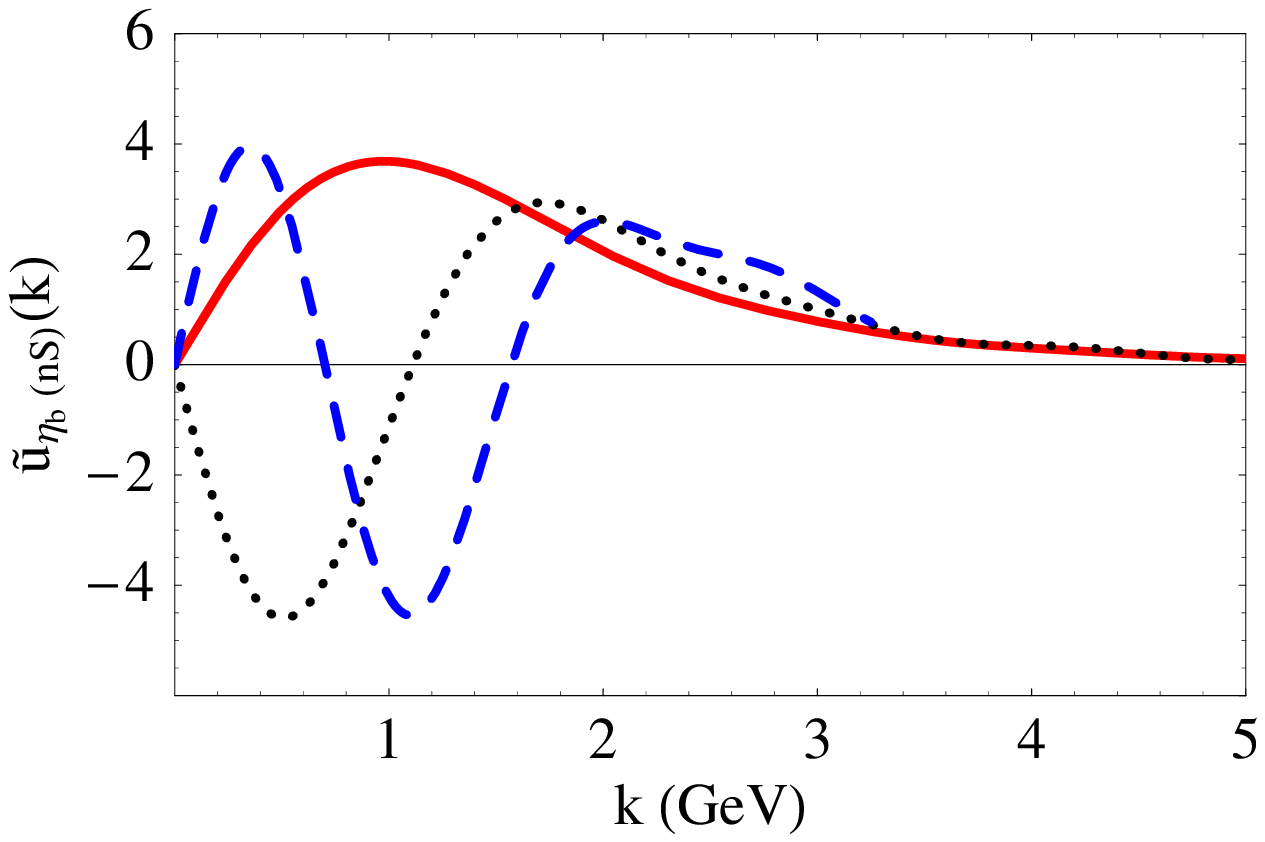}
}
\subfigure{
\includegraphics[scale=0.49]{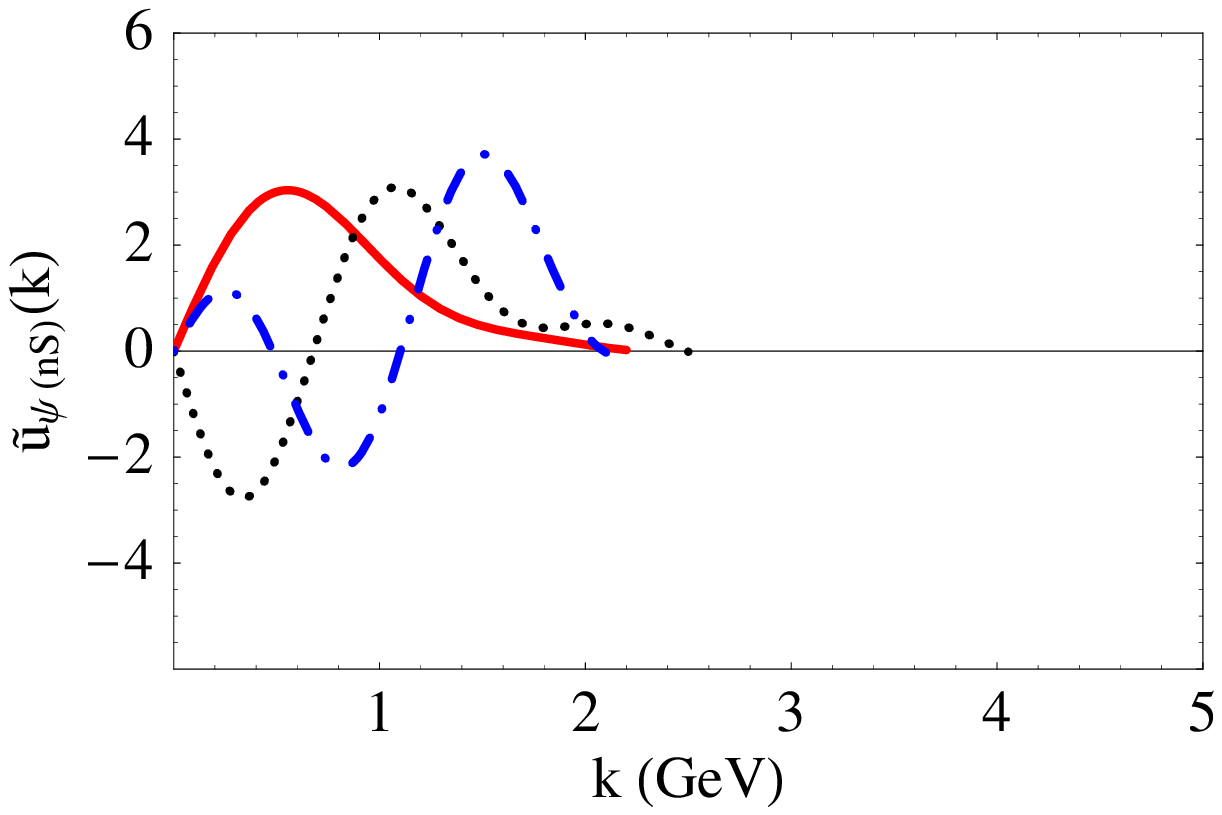}
}
\hspace*{1cm}
\subfigure{
\includegraphics[scale=0.52]{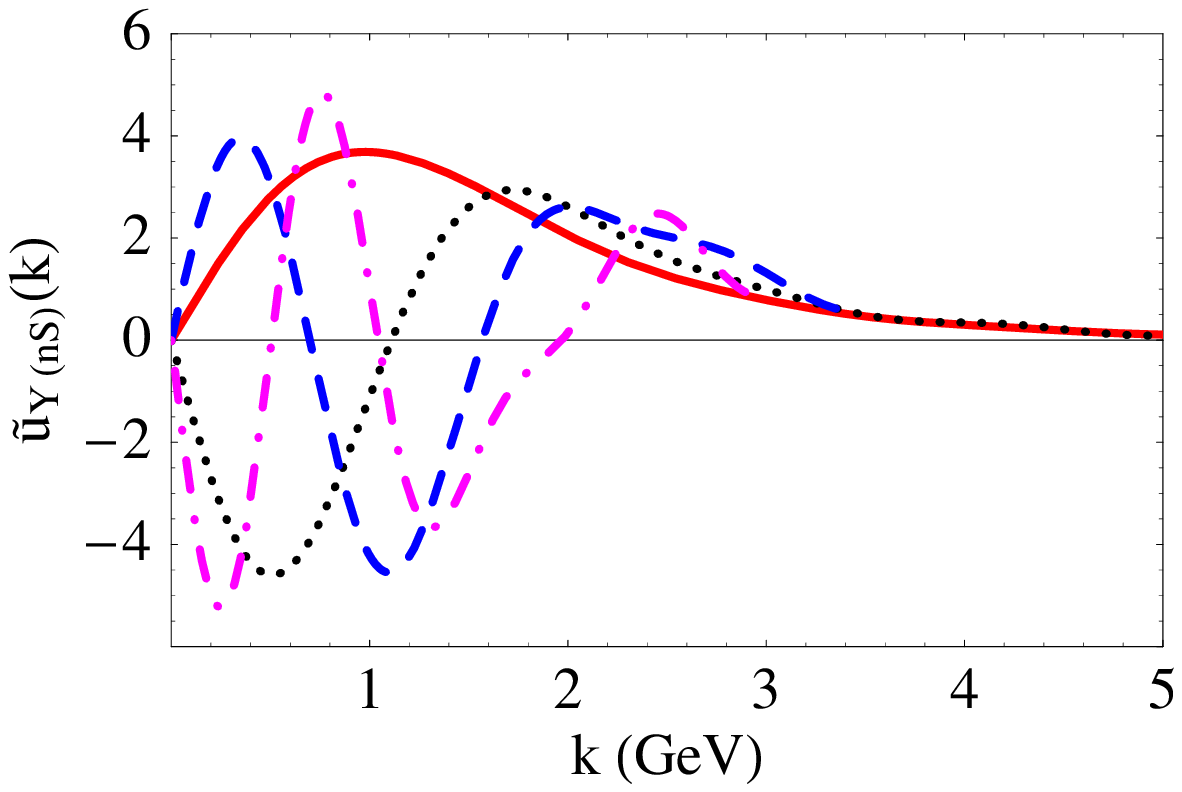}
}
\caption{The momentum wave functions of the first three states of $\eta_c$(nS) (top left), $\eta_b$(nS) (top right), $J/\psi$(nS) (bottom left) and $\Upsilon$(nS) (bottom right). The continuos line represents the 1S state,  the
dotted line represents  the 2S state, the dashed line represents  the 3S state, and the dashed-dotted line represents the 4S state. The wave functions are dimensionless: they are normalized as $\int dk \, |\tilde{u}(k)|^2=2M$. }
\label{wfunc}
\end{figure}

The obtained spectrum and decay constants of $c\bar{c}$ and  $b\bar{b}$ S-wave mesons are collected  in Table \ref{mass};  in Fig. \ref{wfunc}  the corresponding  wave functions are depicted. It is interesting to notice that $f_{\eta_c}$ turns out to be compatible with a determination obtained by the CLEO Collaboration: $f_{\eta_c}=335\pm75$ MeV \cite{Edwards:2000bb}.

\begin{table}[h]
\caption{Masses of pseudoscalar and vector $c\bar{c}$ and $b\bar{b}$ states compared to the experimental data. In the third column the decay constants, computed using (\ref{decayconst}), are reported.}\label{mass}
\begin{center}
\begin{tabular}{|c|c|c||c|c|c|}
\hline
\, Particle \, & Th. mass (MeV)& Exp. mass (MeV)  \cite{pdg} & Decay const. (MeV) \\
\hline
$\eta_c$ & 3025.3 & 2980.3 $\pm$ 1.2 & 342 \\
\hline
$\eta'_c$ & 3603.5 & 3637.0 $\pm$ 4& 266 \\
\hline
$\eta''_c$ & 4039.3 & &195\\
\hline
$J/\psi$ & 3079.8 & 3096.916 $\pm$ 0.011 & 356\\
\hline
$\psi'$ & 3624.3 & 3686.09 $\pm$ 0.04 & 237 \\
\hline
$\psi''$ & 4057.0 & 4039$\pm$1 & 185 \\
\hline
\hline
$\eta_b$ & 9433.9 & 9388.9 $^{+3.1}_{-2.3}$ (stat) $\pm$ 2.7 (syst) \cite{:2008vj} & 637 \\
\hline
 $\eta'_b$ & 9996.8 & & 430 \\
\hline
 $\eta''_b $& 10347.5 &  & 367 \\
 \hline
 $\Upsilon$ & 9438.3 &9460.30$\pm$0.26 & 686 \\
 \hline
 $\Upsilon(2S)$ & 9998.6 & 10023.26 $\pm$ 0.31 & 484 \\
  \hline
 $\Upsilon(3S)$ & 10348.8 & 10355.2 $\pm$0.5 & 335 \\
  \hline
 $\Upsilon(4S)$ & 10622.3 & 10579.4 $\pm$1.2 & 301 \\
\hline
\end{tabular}
\end{center}
\label{default}
\end{table}%

Using the computed values of $f_P$ and $f_V$, it is possible to determine the widths $\Gamma_{\gamma\gamma}$ of the radiative decays $\eta_{b,c}(nS)\rightarrow \gamma\gamma$, and the  widths $\Gamma_{\ell^+\ell^-}$ of the processes $\psi(nS)\rightarrow \ell^+\ell^-$ and $\Upsilon(nS)\rightarrow \ell^+\ell^-$. The widths can be easily computed using the  effective Lagrangians \cite{ Lansberg:2006dw, Lansberg:2006sy}:
\begin{eqnarray}\label{efflag}
\mathcal{L}^{\gamma\gamma}_{eff}&=&-i\, c_1 (\bar{q}\, \gamma^\sigma \gamma^5 q)\epsilon_{\mu\nu\rho\sigma} F^{\mu\nu} A^\rho  \nonumber\\
\mathcal{L}^{\ell\bar{\ell}}_{eff}&=&- c_2 (\bar{q}\, \gamma^\mu q) (\ell\gamma_\mu \bar{\ell}) 
\end{eqnarray}
where
\begin{eqnarray}
c_1&=&\frac{Q^2 \, 4 \pi \, \alpha_{em}}{(M^2+E_b M)} \nonumber\\
c_2&=&\frac{Q\, 4 \pi \, \alpha_{em}}{M^2} \,.
\end{eqnarray}
One obtains
\begin{eqnarray}\label{gamma}
\Gamma_{\gamma\gamma}&=&\frac{4\pi \, Q^4 \alpha_{em}^2M^3f_P^2}{(M^2+E_b M)^2} \nonumber\\
\Gamma_{\ell^+\ell^-}&=&\frac{4\pi \, Q^2 \alpha_{em}^2 f_V^2}{3\, M}\,,
\end{eqnarray}
where $Q$ is the electric charge (in units of  $e$) of the constituent quark  and $E_b=2 m-M$ is the binding energy. 

The values obtained for the pseudoscalar mesons are shown in Table \ref{width0}, together with recent theoretical results. The prediction for the $\eta_c$ radiative decay width is compatible with the experimental value within the error; in the case of  $\eta'_c$, the measurement by the CLEO Collaboration \cite{Asner:2003wv} is  smaller (or marginally compatible) than the obtained theoretical prediction  and than in other calculations \cite{Godfrey:1985xj}. 

Concerning the  $b\bar{b}$ pseudoscalar meson, the theoretical models in Table \ref{width0} predict, for the $\eta_b\rightarrow\gamma\gamma$ decay width, values in the range 230-560 eV; the result obtained in this paper points towards small values in this range.  

\begin{table}[b]
\caption{Decay widths $\Gamma_{\gamma\gamma}$ (in KeV) of  pseudoscalar states in two photons. The value denoted by * is reported by the PDG \cite{pdg} as a datum not included in the summary tables.}\label{width0}
\begin{center}
\begin{tabular}{|c|c|c|c|c|c|c|}
\hline
\,Particle\, & This paper & Lansberg {\it et al.} \cite{Lansberg:2006sy} & Lakhina {\it et al.} \cite{Lakhina:2006vg} & Kim {\it et al.} \cite{Kim:2004rz} & Ebert {\it et al.} \cite{Ebert:2003mu} & Exp.  \\
\hline
$\eta_c$ & 4.252 &  7.46 & 7.18&  7.14$\pm$0.95&5.5   &7.2 $\pm$ 0.7 $\pm$ 2.0 *  \\
\hline
$\eta'_c$ & 3.306 &  4.1& 1.71& 4.44$\pm$0.48 & 1.8    &  1.3$\pm$0.6  \cite{Asner:2003wv}  \\
\hline
$\eta''_c $& 1.992 &  & 1.21 && & \\
\hline
\hline
 $\eta_b$ & 0.313 & 0.560 & 0.230 & 0.384$\pm$ 0.047 & 0.350  &  \\
\hline
$\eta'_b$ & 0.151 & 0.269 & 0.070 & 0.191 $\pm$ 0.025 & 0.150 &  \\
\hline
 $\eta''_b $& 0.092 & 0.208 & 0.040 & & 0.100 &  \\
\hline
\end{tabular}
\end{center}
\label{default}
\end{table}%

For vector mesons, the predicted and the experimental values of the leptonic decay widths are reported in Table \ref{width1}. There is an overall agreement, excluding a  discrepancy in the $\Upsilon$(3S) that could be attributed to a possible $D$-wave component in this meson.

\begin{table}[t]
\caption{Decay widths $\Gamma_{\ell^+\ell^-}$ (in KeV) of  vector mesons.}\label{width1}
\begin{center}
\begin{tabular}{|c|c|c|}
\hline
\,Particle\, & This paper & Exp. \cite{pdg} \\
\hline
$J/\psi$ & 4.080& 5.55$\pm$ 0.14$\pm$0.02\\
\hline
$\psi'$ & 2.375& 2.38 $\pm$ 0.04\\
\hline
$\psi''$ & 0.836& 0.86$\pm$0.07\\
\hline
\hline
 $\Upsilon$ & 1.237 & 1.340 $\pm$ 0.018 \\
\hline
$\Upsilon(2S)$ & 0.581 & 0.612 $\pm$ 0.011  \\
\hline
 $\Upsilon(3S)$& 0.270& 0.443$\pm$ 0.008  \\
\hline
$\Upsilon(4S)$& 0.212& 0.272$\pm$ 0.029  \\
\hline

\end{tabular}
\end{center}
\label{default}
\end{table}%

In conclusion,  the decay constants and the radiative decay widths of $b\bar{b}$ and $c\bar{c}$ pseudoscalar mesons, computed within a semirelativistic quark model which uses a potential inspired by the AdS/QCD correspondence, are compatible  with the experimental data, in particular in the case of $f_{\eta_c}$ and $\Gamma_{\gamma\gamma}(\eta_c)$. The measurement of  $\Gamma_{\gamma\gamma}(\eta'_c)$ carried out by the CLEO Collaboration \cite{Asner:2003wv} is not reproduced, since the obtained result  differs by  more than $2\sigma$. In this respect, our result follows most  theoretical models  \cite{Lansberg:2006dw, Kim:2004rz, Ebert:2003mu, Lakhina:2006vg, Godfrey:1985xj}, which predict higher values for 
$\Gamma_{\gamma\gamma}(\eta'_c)$, although in some cases within the experimental error. This might suggest that  the disagreement could be attributed to the systematics of the experimental measurement, namely, to the assumption  that $\eta_c$ and $\eta'_c$ have the same branching fractions to the final state $K_SK\pi$. As for $\eta_b$, the prediction of the $\eta_b\rightarrow\gamma\gamma$ decay width suggests that this decay mode could be observed in the forthcoming experimental analyses.

\vspace{1cm}

{\bf Acknowledgements}

I would like to thank P.~Colangelo, F.~De Fazio, S.~Nicotri and T.~N.~Pham for useful suggestions and discussions. This work was supported in part by the EU Contract No. MRTN-CT-2006-035482, "FLAVIAnet".


\end{document}